\documentclass[%
 reprint,
 twocolumn,amsmath, amssymb, aps,superscriptaddress,
 prx
]{revtex4-2}

\usepackage{graphicx}
\usepackage{bm}
\usepackage{physics}
\usepackage{xcolor}
\usepackage[normalem]{ulem}
\usepackage{mathrsfs}
\usepackage{tikz}

 \usepackage{cleveref}
\newif\ifstartedinmathmode
\newcommand\encircled[1]{%
  \relax\ifmmode\startedinmathmodetrue\else\startedinmathmodefalse\fi%
  \tikz[baseline,anchor=base]{%
  \node[draw,circle,outer sep=0pt,inner sep=.2ex]
    {\ifstartedinmathmode$#1$\else#1\fi};}%
}

\def\k{{\boldsymbol k}}

\makeindex

\begin{document}

\title{
Observation of pattern stabilization in a driven superfluid
}

\author{Nikolas Liebster}
\email{pattern-formation@matterwave.de}
\affiliation{Kirchhoff-Institut f\"{u}r Physik, Universit\"{a}t Heidelberg, Im Neuenheimer Feld 227, 69120 Heidelberg, Germany}
\author{Marius Sparn}
\affiliation{Kirchhoff-Institut f\"{u}r Physik, Universit\"{a}t Heidelberg, Im Neuenheimer Feld 227, 69120 Heidelberg, Germany}
\author{Elinor Kath}
\affiliation{Kirchhoff-Institut f\"{u}r Physik, Universit\"{a}t Heidelberg, Im Neuenheimer Feld 227, 69120 Heidelberg, Germany}
\author{Jelte Duchene}
\affiliation{Kirchhoff-Institut f\"{u}r Physik, Universit\"{a}t Heidelberg, Im Neuenheimer Feld 227, 69120 Heidelberg, Germany}
\author{Keisuke Fujii}
\affiliation{Institut f\"{u}r Theoretische Physik, Universit\"{a}t Heidelberg, Philosophenweg 19, 69120 Heidelberg, Germany}
\author{Sarah L. G\"{o}rlitz}
\affiliation{Institut f\"{u}r Theoretische Physik, Universit\"{a}t Heidelberg, Philosophenweg 19, 69120 Heidelberg, Germany}
\author{Tilman Enss}
\affiliation{Institut f\"{u}r Theoretische Physik, Universit\"{a}t Heidelberg, Philosophenweg 19, 69120 Heidelberg, Germany}
\author{Helmut Strobel}
\affiliation{Kirchhoff-Institut f\"{u}r Physik, Universit\"{a}t Heidelberg, Im Neuenheimer Feld 227, 69120 Heidelberg, Germany}
\author{Markus K. Oberthaler}
\affiliation{Kirchhoff-Institut f\"{u}r Physik, Universit\"{a}t Heidelberg, Im Neuenheimer Feld 227, 69120 Heidelberg, Germany}

\begin{abstract}
The formation of patterns in driven systems has been studied extensively, and their emergence can be connected to a fine balance of instabilities and stabilization mechanisms. 
While the early phase of pattern formation can be understood on the basis of linear stability analyses, the long-time dynamics can only be described by accounting for the interactions between the excitations generated by the drive.
Here, we observe the stabilization of square patterns in an interaction-driven, two-dimensional Bose-Einstein condensate. 
These patterns emerge due to inherent high-order processes that become relevant in the regime of large phonon occupations.
Theoretically, this can be understood as the emergence of a stable fixed point of coupled nonlinear amplitude equations, which include phonon-phonon interactions. 
We experimentally probe the predicted flows towards such a stable fixed-point, as well as repulsion from a saddle fixed-point, using the experimental control unique to quantum gases.
\end{abstract}

\maketitle

\section{Introduction} Many physical systems exhibit the formation of large scale patterns, both in the ground state or in the course of dynamics. 
Understanding these structures can provide simplification as well as classification of complex phenomena. 
These patterns range from simple structures such as stripes and lattices to more complex arrangements such as spirals, and emerge in a wide variety of disciplines \cite{CrossHohenberg,ARECCHI19991, Koch1994,Maini1997}. 
In the last decades, powerful theoretical tools to analyze and classify pattern formation have been developed.
In certain cases, the development  and stability of patterns can be analyzed by calculating the evolution of amplitudes of elementary patterned states such as stripes, i.e., $A_j e^{i\mathbf{k}_j \cdot \mathbf{x}} + c.c.$.
The complex amplitudes $A_j$ are assumed to vary slowly in time and are used to capture the dynamics resulting from complex underlying details.
In two dimensions, under the assumptions of rotational invariance as well as generalized parity, translation, and inversion symmetry, one constructs the phenomenological amplitude equation \cite{CrossHohenberg,cross2009pattern}
\begin{equation}
\label{eq:amplitudeeq}
   \frac{\mathrm{d}A_j}{\mathrm{d}t}=\varepsilon A_j-\kappa\left|A_j\right|^2A_j+\kappa \sum_{i \neq j} G\left(\theta_{i j}\right)\left|A_i\right|^2 A_j.
\end{equation}
Here, $\varepsilon$ describes exponential growth of the individual stripes, $\kappa$ the self-interaction of a single stripe pattern, and  $\kappa\,G(\theta_{i j})$ the cross-interaction between stripes.
The first nonlinear term describes self-interaction leading to saturation, while the second is the interaction between different stripes with angle $\theta_{ij}$ between them.
While $|\mathbf{k}_j| = k_c$ is assumed to be constant, the direction is not fixed.
The functional form of $G(\theta_{i j})$ determines the stability of certain arrangements, for example square lattices.
The mathematical form of this equation is generic and does not depend on microscopic details. 
However, stabilization dynamics depend on system-specific nonlinearities and the microscopic description.

\begin{figure*}[htb]
\centering
\includegraphics{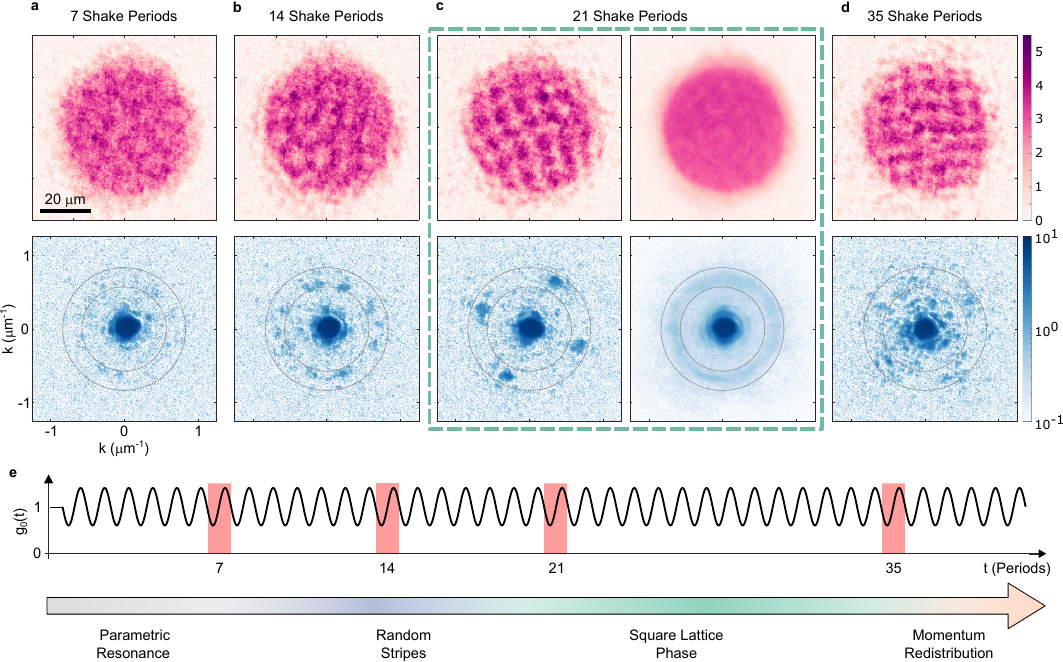}
\caption{\textbf{Structure Formation.} \textbf{a}, Real space (top row) and momentum space (bottom row) distributions after 7 shake periods with $r = 0.4$ and $\omega_d = 2\pi \times 400\mathrm{Hz}$. The dashed lines in momentum space show the region of resonant momenta, centered on $k_c = 2\pi \times 0.11$\,µm$^{-1}$. \textbf{b}, Distributions after 14 periods, where density modulations have become more apparent in real space and appear in momentum space as back-to-back correlated, randomly oriented peaks.  \textbf{c}, Structures in the square lattice phase. The left column shows single realizations, whereas the right column shows averaged real and momentum space distributions. The smooth mean distributions indicate that structures are formed spontaneously in random directions. \textbf{d}, Late times show that square lattices at the characteristic length scale are still apparent, but other momenta also become occupied. Both color codes indicate the signal in atoms per pixel. \textbf{e}, The interaction is periodically modulated with a single frequency and a non-zero offset. The colored arrow indicates the various phases during time evolution.}
\label{fig:Fig1}
\end{figure*}

In driven classical fluids, experimental and theoretical works have investigated pattern formation, including geometry-selective nonlinearities in a variety of parameter regimes \cite{CrossHohenberg,Milner_1991,ZHANG_1997}.
Phenomenologically similar dynamics can occur in superfluids, realized using ultracold Bose gases.
The drive leads to the exponential growth of density waves ($\varepsilon > 0$) at a characteristic wavenumber $k_c$. 
The corresponding length scale is set by the driving frequency through the dispersion relation of elementary excitations to good approximation, and a mean background interaction leads to a non-vanishing $\kappa$.

In superfluids, the role of high-order non-linear processes in the stabilization of specific pattern geometries has not been studied.
In one dimension, Faraday instabilities have been demonstrated  \cite{Kevrekidis:2004,Engels2007_FaradayinBEC, Nicolin:2007, Capuzzi:2008,Tang:2011,Nicolin2011,Westbrook2012_DynaCasi, Smits2018_SpaceTimeCrystal, Jason2019,HernandezRajkov_2021,Dupont2023}, and saturation of density wave contrast was observed \cite{Engels2007_FaradayinBEC,Smits_2020,Dupont2023}.
In two dimensions, square and hexagonal patterns were produced using two drive frequencies at given ratios, which lead to geometry-selective scattering processes between energy shells \cite{Zhang2020}.
Alternatively, driving at  frequencies resonant to surface modes in a harmonic trapping potential has been shown to produce highly-structured surface waves \cite{Kwon2021}, rather than structures that emerge from the bulk.

Here, we report on the spontaneous emergence of a square lattice pattern in a two-dimensional BEC, driven with a single frequency.
The pattern is a result of inherent high-order non-linear processes of the system, which become relevant in the regime of large occupations of phonons.
The state is stabilized due to a cubic nonlinearity similar to \cref{eq:amplitudeeq} and emerges naturally as a result of two factors: large occupations of produced excitations and a positive background interaction, which ensures significant coupling between phonons.
We show that the stability of the square lattice can be understood in terms of a stable fixed point of amplitude equations, whereas lattice solutions  with $\theta_{1,2}$ very different from 90° are unstable, and use state of the art control of our system to experimentally probe the theoretical framework.
The observed dynamics reveal the stability and instability landscapes underlying the dynamical system.

\begin{figure*}[htb]
\centering
\includegraphics{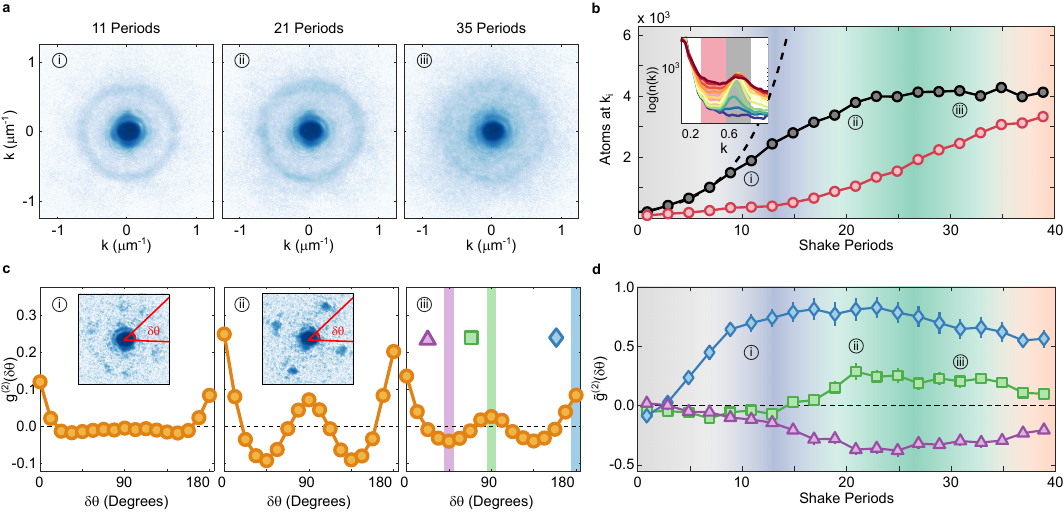} 
\caption{\textbf{Growth and Stability of Structures.} \textbf{a}, Mean momentum distributions at three representative drive times. Excitations at the resonant momentum are pronounced, and show radial symmetry. The occupations at $k_c$ saturate, and at later times non-resonant momenta also become populated.
\textbf{b}, A quantitative analysis of occupations in momentum space. The black data points show occupations at the resonant momentum $k_c$, whereas the red points indicate off-resonant momenta. The inset shows the occupations at all momenta on a log scale from early (blue) to late (red) times; shaded regions correspond to the area yielding black and red data points.
\textbf{c}, Correlation functions $g^{(2)}(\delta \theta)$ at the same drive times as the momentum distributions plotted in (\textbf{a}). Early times show back-to-back correlations but no other structure. In the square lattice phase, peaks at $\delta \theta = 90^\mathrm{o}$ appear, as well as anti-correlations at other angles. Later times show dampened correlations but retain the peak at $\delta \theta = 90^\mathrm{o}$. Insets show representative single realizations at the corresponding drive times.
 \textbf{d}, Extracted values of $\tilde{g}^{(2)}(\delta \theta)$ for $\delta \theta = 45^\mathrm{o}$ (purple), $\delta \theta = 90^\mathrm{o}$ (green), and $\delta \theta = 180^\mathrm{o}$ (blue). Near saturation of momentum occupations (\textbf{b}), the system begins to form $90^\mathrm{o}$ correlations, while $45^\mathrm{o}$ correlations are suppressed.
 The initial increase of back-to-back correlations at early times is an artifact due to enhanced signal-to-background. 
 For all plots, standard errors are either shown or are smaller than the markers.}
\label{fig:Fig2}
\end{figure*}

As is generically the case in systems with spontaneously broken translational symmetry, boundary effects play an important role in the development and stability of emergent patterns; we implement a novel box potential with slanted walls.
The resulting trap suppresses coherent reflections of quasiparticles at the boundary while conserving atom number and leads to dynamics similar to an infinitely extended system.
The effective absorption results from a combination of slowing of the wavefront due to the gradual decrease of the density at the edge, as well as roughness of the potential, which both lead to scrambling of the reflected wavefronts.
This makes it possible to experimentally demonstrate the emergence of square lattices, which are explained theoretically in the infinitely extended limit (see Supplemental Materials).
In the following, we first present experimental results on the emergence of these patterns in our driven condensate, and then detail the theoretical description of the non-equilibrium fixed point.

\section{Emergence of Lattices in Experiment} 
We experimentally realize a quasi two-dimensional BEC of around 30,000 $^{39}$K atoms, with trapping frequency $\omega_z = 2\pi \times 1.5\,\mathrm{kHz}$ in the gravity direction.
In the horizontal plane, the trap shape is circularly symmetric and flat, with walls that are linearly slanted, such that the density distribution of the cloud is uniform and falls linearly at the edges (see Supplemental Materials). 
This is crucial for the emergence of square patterns, as this density distribution implements absorptive boundaries. 
To achieve gain at a specific length scale ($\varepsilon>0$), we periodically change the s-wave scattering length $a_s$ with a frequency $\omega_d$ by varying the external magnetic field near the $561$\,G Feshbach resonance \cite{HadziFeshbach}, such that $a_s(t) = \Bar{a}_s(1 - r \sin \omega_d t)$, with $0 < r < 1$ and $\Bar{a}_s$ set to $100 a_0$, with $a_0$ the Bohr radius. 
The resulting chemical potential is $\mu \sim 2\pi \times 300\,$Hz.

Single shots of \textit{in situ} density distributions as well as momentum distributions after various drive periods are shown in \cref{fig:Fig1}. 
Density distributions are imaged directly after the drive with high field absorption imaging \cite{Hans2021}. 
Alternatively, we extract the momentum space distribution with a phase space rotation, by rapidly switching off the atomic interaction using the zero-crossing of the Feshbach resonance and allowing the cloud to propagate in a weak harmonic trap in the horizontal plane for a quarter period ($2 \pi \times 4.7\,$Hz) \cite{MomentumSpaceCornell,Murthy2014}. The vertical confinement is ramped down to a weak value, just enough to retain the atoms within the focal plane of the imaging system.

At early times (7 periods of driving, \cref{fig:Fig1}a), excitations appear at the critical length scale $2\pi / k_c$ but no obvious pattern has yet emerged.
In momentum space, the excitations appear distributed on a ring of the resonant momentum $k_c$.
These early times are well understood to show growth due to parametric resonance \cite{Faraday:1837,Staliunas2002,Carusotto2010,Busch2014}, as the energy input from the drive produces pairs of quasi-particles, where the frequency of each quasi-particle $E$ is half a drive quantum, $E = \omega_d/2$. 
The corresponding $k_c$ is determined by the Bogoliubov dispersion relation, as has been previously predicted and measured in a variety of experiments \cite{Staliunas2002,Zhang2020}.

Continuing the drive (14 periods, \cref{fig:Fig1}b), the density distribution shows enhanced modulation at the critical length scale, but still shows no global structure. 
This regime can be understood as the coexistence of different, randomly oriented stripe patterns.
This is directly revealed by the momentum distributions, which show clear back-to-back correlated peaks at the resonant momentum.

At intermediate times, global square lattice patterns emerge (21 periods, \cref{fig:Fig1}c). 
This can be observed directly in real space as a square density modulation, or in momentum space as four momentum peaks, each separated by $90^\mathrm{o}$. 
Averaging all realizations in real and momentum space, we recover the homogeneous density distribution and a radially symmetric ring in momentum space (second column of \cref{fig:Fig1}c). 
This confirms that square patterns are formed spontaneously, i.e., with random orientation and spatial phase.
For late times, real space distributions still show some structure, while momentum space distributions reveal that many other momenta beside $k_c$ have become occupied, indicating that the model of superimposed stripes at a specific momentum as in \cref{eq:amplitudeeq} is insufficient to capture the late-time dynamics.

The emergence and persistence of square lattices can be quantified by analyzing occupations as well as angular correlations in momentum space. 
In \cref{fig:Fig2}a, momentum distributions averaged over many realizations are shown. 
The ring at $k_c$ is pronounced and saturates in amplitude, and at late times a broad range of off-resonant momenta becomes populated.
Figure \ref{fig:Fig2}b shows the occupation increase relative to the unperturbed condensate, centered at $k_c = 2\pi \times 0.11\,\mu\mathrm{m}^{-1}$  as well as off-resonant momenta centered at $k = 2\pi \times 0.07\,\mu\mathrm{m}^{-1}$, integrated over a width of $2\pi\times0.04\,\mu\mathrm{m}^{-1}$.
Early times show exponential growth at $k_c$, as expected by parametric resonance  (dashed black line in \cref{fig:Fig2}b). 
After around 10 periods, a deviation from this exponential growth due to saturation becomes clear, and eventually occupations reach a steady value of 4,000 atoms in the summed region, relative to 30,000 condensed atoms. 
When resonant momenta $k_c$ are saturated, non-resonant momenta also begin to grow and eventually have similar occupations to the resonant momenta (red line in \cref{fig:Fig2}b). 
We will show in the following that in the regime where occupations are saturated but modes with $|\mathbf{k}| = k_c$ dominate, a description similar to \cref{eq:amplitudeeq} captures the dynamics.

To describe the spatial structure of density waves, we analyze the correlations of particle distributions along the ring at $k_c$ in single shots.
Momentum distributions of single realizations are analyzed by binning occupations along the resonant momentum as a function of the polar angle $\theta$. 
The auto-correlation of this vector is calculated, and these correlations are averaged over all realizations, yielding
\begin{equation}
    g^{(2)}(\delta \theta) = \left< \frac{\left <n(k_c,\theta) \, n(k_c,\theta + \delta \theta)\right >_{\theta}}{\left <n(k_c,\theta)\right >_{\theta}^2} - 1\right>.
    \label{eq:g2}
\end{equation}
Here, $n(k,\theta)$ is the experimentally measured momentum distribution, and the average over $\theta$ indicates the average over absolute orientation while keeping relative angle $\delta \theta$ fixed, whereas the average of the whole expression indicates an average over all shots of a given data set.
We further define $\tilde{g}^{(2)}(\delta \theta) = g^{(2)}(\delta \theta)/g^{(2)}(0)$, a normalized value of the correlation that accounts for variations in total signal on the resonant ring. In this normalization, $\tilde{g}^{(2)}(\delta \theta) = 1$ indicates that every bin at $\theta$ has the same magnitude as the bin at $\theta + \delta \theta$, regardless of total occupations. 
Note that because this is an auto-correlation, the function only contains information about angle differences, $\delta\theta \in [0^\mathrm{o},180^\mathrm{o}]$.

\begin{figure*}[t]
\centering
\includegraphics{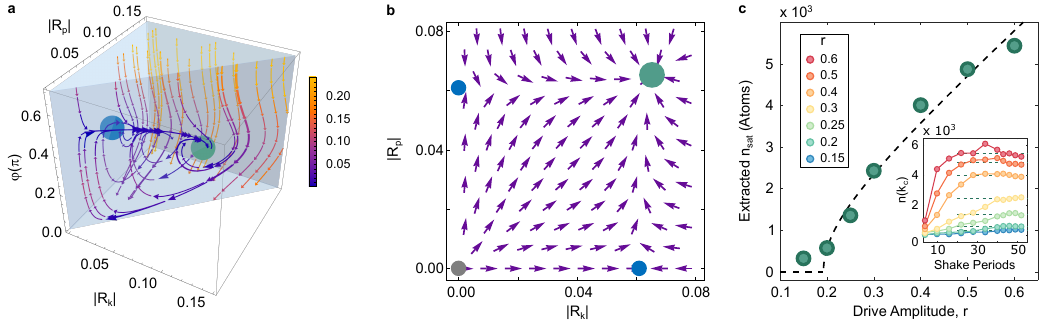}
\caption{\textbf{Attractive Fixed Point.} \textbf{a} Flow diagram of a 3D cut from the full 4D phase space of the dynamics of $R_{k/p}$, where $\theta = 90^{\mathrm{o}}$, $\varphi_k = \varphi_p$, $r = 0.5$ and $\Gamma = 0.4\alpha$.
Arrows show flow lines from the projection of the first derivative $\frac{d}{d t}R_{k/p}$ onto this cut, and colors show the flow speed. 
Two in-spirals towards fixed points (blue and green points) are apparent.
\textbf{b}, Stability diagram for $\delta \theta = 90^{\mathrm{o}}$. Arrows indicate the direction of the second derivative $\frac{d^2}{d t^2}R_{k/p}$ in the plane where $\varphi_k = \varphi_p = \pi/4$ for $\Gamma = 0$.
The three unique fixed points are shown as colored points: unstable points for the homogeneous (gray) and stripe (blue) solutions, and a stable point for a lattice solution (green). 
\textbf{c}, Extracted values of saturated momentum occupations as a function of driving amplitude $r$. 
The inset shows occupations near $k_c$ for different values of $r$, integrated over the same region as in \cref{fig:Fig2}b.
Saturated occupations $n_{\mathrm{sat}}$ are determined by averaging the last 5 points, and the value is shown as the dashed line in the inset and as green points in the main plot.
The green points of the main plot approximate the occupations at the fixed point at various drive amplitudes. The dashed line is a fit to the data with the theoretically predicted functional form  $F(r) = a\sqrt{(r/b)^2-1}$.
Standard errors are smaller than the data points.}
\label{fig:Fig3}
\end{figure*}

Exemplary $g^{(2)}(\delta \theta)$ are plotted in \cref{fig:Fig2}c.
After 11 drive periods (\cref{fig:Fig2}c, \encircled{i}), an auto-correlation peak as well as a peak at $\sim\!180^\mathrm{o}$ appear, but no other structure is apparent, indicating randomly oriented stripes.
During the square lattice phase (\cref{fig:Fig2}c, \encircled{ii}), these correlations show a clear peak at $90 ^\mathrm{o}$ and $180 ^\mathrm{o}$, while angles between these peaks become anti-correlated. 
At even later times when redistribution effects begin to dominate (\cref{fig:Fig2}b, \encircled{iii}), one can see that correlations have a slightly lower contrast, but the characteristic peaks at  $90 ^\mathrm{o}$ and $180 ^\mathrm{o}$ remain. 
While the positive correlation at $90 ^\mathrm{o}$ indicates that square lattices are produced frequently, the anti-correlation of angles other than $90 ^\mathrm{o}$ indicates that single square lattice structures over the whole system are dominant in single realizations, as negative correlations are the result of suppressed signal relative to the average value.

The dynamics can be characterized by extracting specific values of these correlations as a function of drive time.
Figure \ref{fig:Fig2}d shows extracted values of correlation functions for $\delta \theta = 45 ^\mathrm{o}$, $90 ^\mathrm{o}$, and $180^\mathrm{o}$. 
Correlations corresponding to the lattice structure emerge in the course of the dynamics, when occupation numbers and back-to-back correlations are fully established, indicating that lattice formation is a higher-order process.
These correlation values persist over a number of periods, indicating the emergence of the steady-state, until very late times where redistribution effects dampen correlations at all angles.
We note that lattices emerge over a large range of experimental parameters, including different background interaction strengths, drive frequencies, and geometries, including a harmonic trapping potential, confirming the robustness of the formation process.

\section{Amplitude Equation}
The emergence of this patterned state can be described by amplitude equations for standing waves on a BEC. 
The evolution of the BEC order parameter $\Psi(\mathbf{x},t)$ is described by the time-dependent Gross-Pitaevskii equation (GPE),
\begin{equation}
    \label{eq:GPE}
    \begin{aligned}
    i \hbar \frac{\partial \Psi(\mathbf{x},t)}{\partial t} =& \Bigg( -\frac{\hbar^2 \nabla ^2}{2 m} + V(\mathbf{x})\\ 
    & + g_0\left(1 - r \sin \omega_d t \right) |\Psi(\mathbf{x},t)|^2 \Bigg) \Psi(\mathbf{x},t).
    \end{aligned}
\end{equation}
Here, $m$ is the atomic mass, $\hbar$ is the reduced Planck's constant, and the interaction strength given by $g_0 = \frac{\sqrt{8 \pi} \hbar^2}{m} \frac{\bar{a}_s}{l_z}$, where  $l_z = \sqrt{\frac{\hbar}{m \omega_z}}$. 
We describe the emerging pattern as a sum of two stripe patterns
\begin{equation}
    \label{eq:ansatz}
     \Psi(\mathbf{x},t) = \Psi_{\text{uni}}(t) \Big[1 + \phi_k(t) \cos{(\mathbf{k}\cdot\mathbf{x})} + \phi_p(t) \cos{(\mathbf{p}\cdot\mathbf{x})} \Big]
\end{equation}
and insert this into the time-dependent GPE, \cref{eq:GPE}, with $V(\mathbf{x}) = 0$. 
$ \Psi_{\text{uni}}(t)$ is a uniform, infinitely extended background field with time evolution $\Psi_{\text{uni}}(t) = \sqrt{n_0}\, \exp[-i\mu t - i (\mu / \omega_{d}) r \cos{\omega t}]$, where $n_0$ is the 2D density. The vectors $\textbf{k}$ and $\textbf{p}$ with $|\textbf{p}| = |\textbf{k}| = k_c$ have an angle $\theta \in [0^\mathrm{o},180^\mathrm{o}]$ between them.
The ansatz \cref{eq:ansatz} extends previous theoretical work, which analyzed the stability of a single standing wave, and did not consider coupling between multiple waves \cite{Staliunas2002}.
Because the excited modes are Bogoliubov quasi-particles, the amplitudes of the standing waves are parameterized by  
\begin{equation}
    \label{eq:Phonon}
    \begin{aligned}
    \phi_{k / p}(t) =&\left(1-\frac{\epsilon+2 \mu}{E}\right) R_{k / p}(t) e^{i \frac{\omega_d}{2} t}\\
    &+\left(1+\frac{\epsilon+2 \mu}{E}\right) R_{k / p}^*(t) e^{-i \frac{\omega_d}{2} t}.
    \end{aligned}
\end{equation}
Here, $E = \sqrt{\epsilon ( \epsilon + 2 \mu)}$ with $\epsilon = \frac{\hbar k_c^2}{2 m}$ is the corresponding Bogoliubov energy in units of frequency. 
$R_{k/p}$ are complex amplitudes that vary slowly in time. 
The phase of $R_{k/p}$ can be understood as a relative phase between the oscillation of the phonon and the drive, and $|R_{k/p}|^2$ corresponds to phonon occupation numbers.

The dynamics of these amplitudes can be understood using a Ginzburg-Landau-type equation \cite{cross2009pattern}.
To derive this equation from \cref{eq:GPE} and \cref{eq:ansatz}, we perform a multiple-timescale analysis \cite{Keisuke}, resulting in the amplitude equations
\begin{equation}
    \label{eq:GinzburgLandau}
    \begin{aligned}
    i \frac{d}{d t} R_k(t)=&-i \alpha R_k^*(t) - i\Gamma R_k(t) + \Delta R_k(t)\\
    &+\lambda\left|R_k(t)\right|^2 R_k(t) \\
    & +\lambda\Big[c_1(\theta)\left|R_p(t)\right|^2 R_k(t)+c_2(\theta) R_p(t)^2 R_k^*(t)\Big]
    \end{aligned}
\end{equation}
and analogously for $R_p$, with the $k$ and $p$ labels exchanged.
Here, $\alpha = r \frac{\mu\epsilon}{2 E}$ is the exponential growth rate from parametric resonance, $\Gamma$ is a phenomenological damping constant, and $\Delta = \frac{\omega_d}{2} - E$ is a detuning term that is set to zero. 
Other constants $\lambda = \mu \frac{5\epsilon + 3\mu}{E}$, $c_1(\theta)$ and $c_2(\theta)$ are set by $k_c$ and the angle between $\mathbf{k}$ and $\mathbf{p}$ (see Supplemental Materials). 
This equation exhibits a similar structure to \cref{eq:amplitudeeq}. 
Exponential growth at early times is given by the difference between gain $\alpha$ and loss $\Gamma$, whereas saturation and coupling of waves results from the nonlinear terms.
Saturation occurs even for a single wave (i.e., $R_k > 0$ while $R_p = 0$), as growth of one stripe is limited due to a GPE-type nonlinear interaction. 
Additionally, the last two terms capture the angle-dependent coupling between $R_k$ and $R_p$ and are characterized by the real functions $c_1(\theta)$ and $c_2(\theta)$. 
These determine whether stripe or lattice solutions are stable at a given angle, as discussed in the generic amplitude \cref{eq:amplitudeeq} with factor $G(\theta)$.
The dynamics shown in Fig. 2 can now be understood in terms of the amplitude equation. Early times show exponential growth of occupations (linear terms). 
Later, saturation occurs as the occupations become large and nonlinear terms become relevant, eventually leading to square patterns.

\begin{figure*}[htb]
\centering
\includegraphics{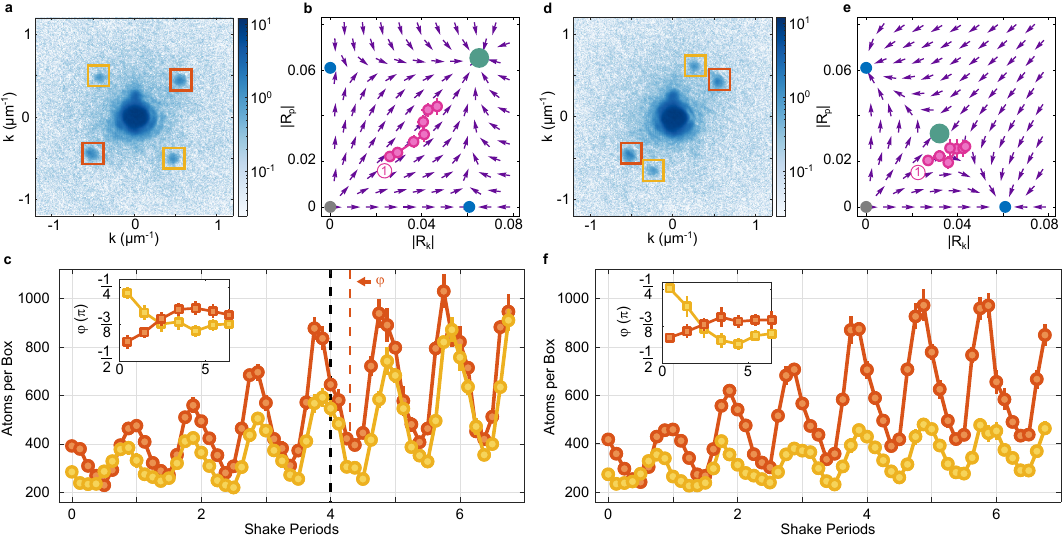}
\caption{\textbf{Dynamics of imprinted lattices} \textbf{a}, Ensemble average of  the initial condition for imprinted patterns at 90$^\mathrm{o}$. Occupations of $R_k$ ($R_p$) are extracted by summing over the red (yellow) boxes. 
\textbf{b}, Extracted $|R|$ values (pink dots) in the theoretically predicted stability diagram for $\theta = 90^\mathrm{o}$.  \textbf{c} Oscillations of average occupations per summation box in momentum space of the imprinted waves, with error bars indicating standard errors. The inset shows the extracted phases of $R_{k/p}$, with 1$\sigma$ fit errors. Though waves are imprinted with different phases, they quickly phase-lock with a slight delay to the drive. \textbf{d} Average momentum distributions for an identically imprinted wave to (\textbf{a}), but with $\theta = 30^\mathrm{o}$. \textbf{e}, Extracted $|R|$ values and the stability diagram for $\theta = 30^\mathrm{o}$, indicating an unstable lattice fixed point. \textbf{f}, Occupations of momenta for the triangular lattice, $\theta = 30^\mathrm{o}$. Despite identical initial conditions to the imprinted wave at  $\theta = 90^\mathrm{o}$, growth of the second stripe is significantly suppressed.}
\label{fig:Fig4}
\end{figure*}

For a set angle $\theta$, the time evolution of the complex amplitudes $R_k$ and $R_p$ takes place in a four-dimensional phase space, as each amplitude influences the magnitude and phase of the other (see Supplemental Materials for coupling between magnitude and phase of complex amplitudes).
For the following discussion, we set the angle to $\theta = 90^\mathrm{o}$.

In order to visualize the structure of the amplitude equations, we utilize flow diagrams, which show the direction and magnitude of the first derivative $\frac{d}{d t}R_{k/p}$ at all points in space.
Flows in the four-dimensional phase space are projected onto the three-dimensional cut where the phases $\varphi_{k/p} = \arg{R_{k/p}}$ are equal, and are illustrated in \cref{fig:Fig3}a, where the color map shows the flow speed at a given point. 
The flow patterns reveal in-spirals toward two fixed points at nonzero amplitude and phase, marked by the blue and green spheres.

To further investigate the structure and stability of these fixed points, we focus on the plane $\varphi_k = \varphi_p$ where all the fixed points lie ($\varphi_{k/p} =\pi/4$ for $\Gamma = 0$). 
Within this plane, by plotting the direction of the force field $\frac{d^2}{d t^2}R_{k/p}$ (see \cite{Keisuke}) we obtain the global stability diagram shown in \cref{fig:Fig3}b. 
Three unique fixed points (i.e., steady states of the driven system, plotted as colored dots) can be identified: one where all amplitudes are zero ($R_{k} = R_{p} = 0$), one where only one density wave has a nonzero amplitude ($R_{k} = 0, |R_{p}| > 0$ or vice versa), and one lattice solution  ($ |R_{k}| = |R_{p}| > 0$). 
The uniform fixed point $R_{k} = R_{p} = 0$ (grey) is unstable in the driven system, and the outgoing arrows represent exponential growth of small non-zero amplitudes (i.e. phonon occupation numbers) due to parametric resonance.
While in one dimension, the single stripe fixed point would be a stable solution, in two dimensions this point is unstable towards forming a lattice at $90^\mathrm{o}$.
We note that the fixed point exists solely due to the nonlinearities of the amplitude equation, and that even in the case of $\Gamma = 0$ the structure and qualitative behavior of the fixed point remains. $\Gamma > 0$ ensures that the fixed point is approached dynamically.

\section{Observation and Probing of the Fixed Point} 
The location of the square lattice fixed point depends on the growth rate $\alpha$, nonlinearity $\lambda$, and damping $\Gamma$. 
While $\alpha$ is a parameter of the linear term that is straightforward to calculate, $\lambda$ describes a nonlinear effect resulting from the interaction between waves, which is theoretically challenging.
The phenomological damping parameter $\Gamma$ must be extracted experimentally. 
The fixed point can be identified by looking at the saturation of occupations in the long-time limit $n_{\mathrm{sat}}$ (see inset of \cref{fig:Fig3}c), where the system is well characterized by the fixed point. 
These are given by $n_{\mathrm{sat}} \propto \frac{1}{\lambda} \sqrt{\alpha^2 - \Gamma^2}$. 
Combining these observations with the extracted growth rate of the occupation number at early times $n(t) = e^{2(\alpha - \Gamma)t}n(t=0)$ (\cref{fig:Fig2}b), we determine these parameters (for details see Supplemental Materials).
We find quantitative agreement with the theoretical prediction for $\alpha$.
The experimentally-determined $\lambda_{\mathrm{eff}}$ is a factor of three larger than the predicted value from the theory framework.
The strongly reduced theoretical model includes neither finite size, occupation-dependent loss nor the presence of additional excitations beyond a two-stripe description.

Building on our experimental capabilities of generating arbitrary potentials and with that density distributions \cite{Viermann2022}, we can test the response of our system to specific angles and phases of structures by explicitly imprinting multiple stripes onto the condensate prior to the drive.
We experimentally seed excitations with amplitudes and phases at values different from the stable fixed point solution.
This is achieved by loading the BEC into a trapping potential with additional periodic spatial modulations to seed stripes.
These periodic potentials are switched off individually with given delay times to set the initial phases of the stripes with respect to the drive.
After the delays, we switch on the driving of the interaction, and detect the populations in the imprinted stripes after a given drive time.
We thus have extensive control over imprinted waves: the angle between waves $\theta$ is set by the shape of the modulated potential, the intensity of the projected light field sets $|R_{k/p}|$, and the time between switching off the modulated potential and starting the drive sets $\varphi_{k/p} = \mathrm{arg}(R_{k/p})$. 

In particular, we contrast the different evolutions of imprinted lattices at two angles, $\theta = 90^\mathrm{o}$ and $\theta = 30^\mathrm{o}$, as shown in \cref{fig:Fig4}.
Initial momentum occupations for the two cases are shown in the left plots of \cref{fig:Fig4}a and d. 
We extract $|\phi_{k/p}|^2$ by summing over a region around the imprinted momenta, indicated by red and yellow boxes, and average the $\pm k$ peaks.
Even for constant phonon occupations, the measured atom numbers in momentum space oscillate in time.
By fitting the function $f(t) = A\left(1-\cos{(\omega_d t + 2\varphi)}\right) + \mathrm{const.}$ to single periods of these oscillating populations, we can extract the slowly-varying magnitude of the stripe amplitudes $|R_{k/p}| = \sqrt{\frac{A\epsilon}{N \mu}}$, and the phase relative to the shaking $\varphi = \varphi_{k/p}$ (see methods).

Imprints for both angles are realized with the same initial occupations and  $\varphi_p(t=0) \sim -\pi/8$, $\varphi_k(t=0) \sim -\pi/2$. 
In both cases, one observes the quick phase-locking of the stripes to the drive, such that $\varphi_p = \varphi_k = -3\pi/8$ (inset), showing that the dynamics of \cref{eq:GinzburgLandau} set not only the occupations but also the phase of phonons.
Despite similar phase evolutions for both angles, the dynamics of the occupations look dramatically different.
While for the square lattice pattern both stripe directions grow over time, in the 30$^{\mathrm{o}}$ case only one of the stripe direction grows.
This can be understood by looking at the stability diagrams for $\theta = 30^{\mathrm{o}}$ and $\theta = 90^{\mathrm{o}}$, plotted in \cref{fig:Fig4}b and e. 
While the lattice solution for $\theta = 90^{\mathrm{o}}$ is a stable fixed point, for $\theta = 30^{\mathrm{o}}$ this point is unstable.
Converting the extracted oscillation amplitudes to phonon amplitudes and plotting these into the respective stability diagrams (pink points in \cref{fig:Fig4}b and e), one sees that amplitudes of waves in the square lattice grow towards the fixed point, while the triangular lattice shows a lack of growth of $R_p$.

These measurements thus give insight into the process of lattice formation at later drive times. 
Once randomly produced stripes have large amplitudes, the interaction between them becomes relevant.
Stripes with orientations close to $\theta = 90^{\mathrm{o}}$ flow towards higher occupation numbers, whereas other orientations flow towards single stripe solutions.
Additionally, these results emphasize the necessity of the soft wall potential, which minimizes reflections of quasiparticles. 
Reflected waves would have a broad distribution of angles relative to the original pattern, and thus cause the deterioration of single square lattices.
Experiments performed with a steep wall trapping potential showed brief emergence of square lattices, but structures are quickly damped and end in disordered patterns.

\section{Outlook}
This work demonstrates the emergence of a stable Faraday pattern in a two-dimensional superfluid, as well as the pattern's stabilization mechanism.
Additionally, we show that the precise control over the experiment enables us to realize highly structured but unstable patterns into the system and observe their dynamics, a capability unique to ultra cold quantum gases.
The stability of the pattern combined with the ability to imprint arbitrary initial configurations opens the possibility to probe additional properties.
As a steady state of a driven superfluid that spontaneously breaks translational symmetry, it could be closely related to supersolidity if distinct sound modes in the lattice and superfluid can be demonstrated.
Control over initial states could be used to probe lattice excitations as well as superfluidity.

\section*{Acknowledgments} We thank M. Hans and C. Viermann for experimental support, and N. Antolini, B. Blakie, L. Chomaz, G. Modugno, S. Stringari, and W. Zwerger for fruitful discussions. This work is supported by the Deutsche Forschungsgemeinschaft (DFG, German Research Foundation) under Germany’s Excellence Strategy EXC 2181/1 - 390900948 (the Heidelberg STRUCTURES Excellence Cluster), under SFB 1225 ISOQUANT - 273811115. This project was funded within the QuantERA II Programme that has received funding from the European Union’s Horizon 2020 research and innovation programme under Grant Agreement No 101017733 as well as the DFG under project number 499183856. N.L. acknowledges support by the Studienstiftung des Deutschen Volkes.

\section*{Appendix A: Experimental System}
The experiment is described in \cite{Viermann2022}. In short, we use a BEC of approximately 30,000 $^{39}$K atoms in the state corresponding to $\ket{F,m_F} = \ket{1,-1}$ at low fields. The trapping frequency in the z-axis is $\omega_z = 2\pi \times 1.5$\,kHz, generated by a blue-detuned lattice. The BEC has a temperature of approximately 20\,nK. The time-averaged chemical potential is $\mu \sim 300$\,Hz for all measurements, which is extracted by scanning driving frequencies, measuring the wavelength of produced density waves, and fitting the Bogoliubov dispersion relation. Approximately 50 realizations are used for measurements in \cref{fig:Fig2}, and 30 in \cref{fig:Fig3} and \cref{fig:Fig4}. The slanted-wall potential is realized with 532$\,$nm light, shaped with a digital micromirror device \cite{Viermann2022} and has the form 
\begin{equation}
V(\mathbf{x}) =\begin{cases}
          0 \quad & \, |\mathbf{x}| < R \\
        \beta (|\mathbf{x}| - R) \quad & \,|\mathbf{x}|\geq R \\
     \end{cases}
\end{equation}
where for our system $R\sim22$\,$\mu$m and $\beta\sim2\pi\times 30\,\mathrm{Hz}/\mu$m.
\\
\\
\section*{Appendix B: Role of Boundaries}
Here, we present a brief study on the role of boundary conditions in the stabilization of patterns in the experiment. 
As described in the main text, slanted walls are a key experimental innovation that enable the experimental observation of stable square lattices, because they significantly reduce the impact of reflections of density waves at the boundaries.
In order to systematically study this phenomenon, we observe the spontaneous emergence of patterns for two trapping geometries: slanted walls and very steep (or hard) walls.

The following experiments use a drive amplitude of $r=0.6$, slightly larger than the amplitude typically discussed in the main text. 
The larger drive amplitude makes dynamics faster and increases the rate of heating, reducing the total lifetime of patterns.
We use a digital micromirror device to program the trapping potential of our cloud to produce not only slanted walls (described above), but also hard walls, whose steepness is limited by the resolution of the optical objective ($\sim$0.5\,µm at 532 nm wavelength).
The hard wall trapping potential has a slightly larger radius such that the mean central density is comparable in both cases. 
Both trapping geometries are loaded identically, and around 50 realizations are used for the following analysis.

\begin{figure*}
\centering
\includegraphics{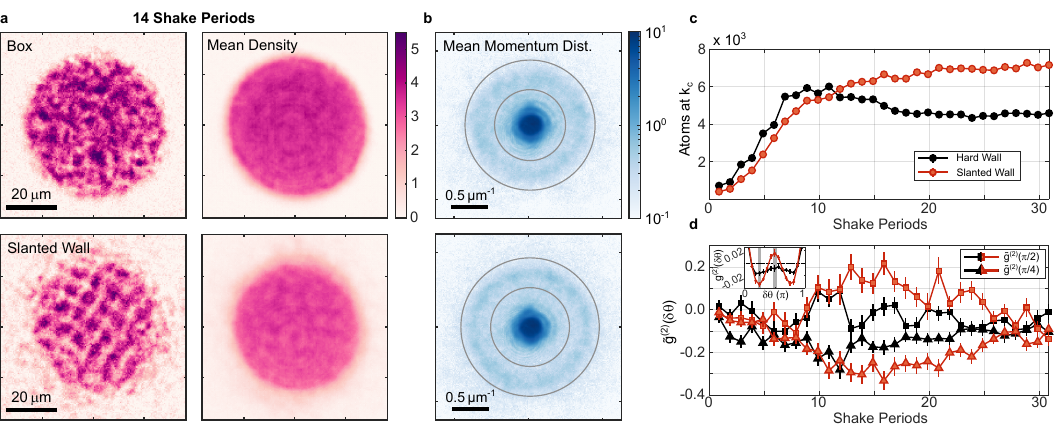}
\caption{\textbf{Effect of Boundaries} a. Single shots and averaged densities for hard (top row) and slanted (bottom) walls. 
b. Averaged momentum distributions show radially symmetric rings, showing the radial symmetry is not broken on average. Despite the seeming disorder in the density distribution of the box trap case, there is no other length scale with significant occupations. The occupations on the ring as well as $g^{(2)}(\delta\theta)$ correlations are calculated in the region indicated by the circles.
c. Quantitative analysis of occupations on ring. Both geometries show exponential growth for early times and saturation for late times.
d. Square pattern correlations. The inset shows $g^{(2)}(\delta\theta)$ correlators after 14 periods. The main plot shows the evolution of $\tilde{g}^{(2)}(\pi/2)$ and $\tilde{g}^{(2)}(\pi/4)$ for the box trap (black) and the slanted wall (red).}
\label{fig:ExtendedDataBoxSlox}
\end{figure*}

Single shots and average densities are shown for both cases in \cref{fig:ExtendedDataBoxSlox}a.
While in the slanted wall case the square lattice patterns are stabilized, in the hard wall case the emergent lattice pattern is quickly distorted, resulting in disordered structures for later times.
As can be seen in the averaged real space and momentum distributions for the hard and slanted walls (\cref{fig:ExtendedDataBoxSlox}b), radial symmetry is not broken on average in either of the traps, as indicated by the homogeneous rings.
Only one dominant length scale is present in both cases, despite the structural disorder in the density distribution of the box trap.

In \cref{fig:ExtendedDataBoxSlox}c, the mean occupations at $k_c$ are plotted, and the pattern correlations are plotted in \cref{fig:ExtendedDataBoxSlox}d. 
For early times, both geometries show exponential growth of occupations and no structure in $g^{(2)}(\delta\theta)$.
After around 8 drive periods, correlations are suppressed at $\delta\theta = \pi/4$ while correlations at $\delta\theta = \pi/2$ grow, indicating the emergence of square lattices in both geometries. 
In the slanted wall trap, the square lattice correlations remain relatively constant; in the box trap, however, correlations are quickly suppressed, showing that the brief emergence of square lattices gives way to disordered stripes.
At the time where square lattices emerge in the slanted wall case (13 periods), the average pattern contrast also becomes larger in the slanted wall trap than in the box trap.
This confirms the theoretical prediction that many competing stripes have lower mean occupations than single patterns \cite{Keisuke}.
Finally, we note that the growth of patterns can be increasingly suppressed by increasing the steepness of the slanted wall, further verifying that reflections play a role in the dynamics in the hard wall case.
\\
\\
\section*{Appendix C: Correspondence between Phonon Amplitudes and Atom Numbers}
In the results described in \cref{fig:Fig4}, we measure occupations of single standing waves, and thus analyze peaks in the experimental momentum space distribution rather than average occupations along the entire resonant momentum ring.
The correspondence between the phonon amplitude $R_k$ and the atom number in a single peak in momentum space $n_k$ is given by the corresponding Fourier amplitude of the \textit{ansatz} given in \cref{eq:ansatz}. 
The observed atom number in a summation box around a momentum peak is given by
\begin{align}
  \label{eq:6}
  n_k = N_0 |R_\k(t)|^2 \Bigl[ 1+2\frac\mu\varepsilon
  \sin^2\left(\frac{\omega_d}{2} t+\varphi_\k(t)\right) \Bigr].
\end{align}
Here, $N_0$ is the condensate atom number, and $\varphi_k$ is the phase of the phonon amplitude $R_k$.
Thus, the number of atoms per momentum peak oscillates at the drive
frequency $\omega_d$.
Other than for data in \cref{fig:Fig4}, measurements of occupations are taken when signal is maximal, which differs by a phase of $\pi$ between real and momentum space.
Period numbers given in the text are rounded to the nearest whole period to avoid confusion.
\\
\\
\section*{Appendix D: Stability Criterion}
\begin{figure}[t]
\centering
\includegraphics{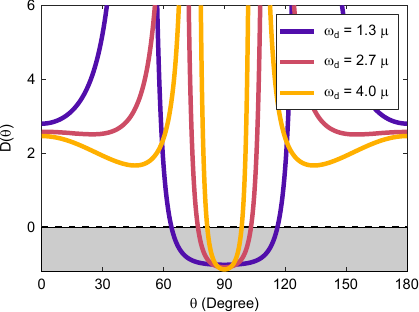}
\caption{\textbf{Instability Curves} The functional form of the instability factor $D(\theta)$. Regions where $D<0$ indicate stable lattice solutions. The divergences are an artifact of scattering processes resonant to the shaking frequency and are discussed in detail in \cite{Keisuke}.}
\label{fig:ExtendedData}
\end{figure}

The angle-dependent prefactors of \cref{eq:GinzburgLandau} are given by \cite{Keisuke}
\begin{widetext}
\begin{equation}
c_1(\theta)=\frac{\mu}{5 \epsilon+3 \mu}\left[4 \frac{\epsilon^2-\mu^2}{\mu \epsilon}+\left(\frac{2 \epsilon+\mu}{\epsilon} \frac{2 \epsilon+\mu}{2 \epsilon \cos ^2 \frac{\theta}{2}+\mu}-\frac{(2 \epsilon-\mu)(\epsilon+2 \mu)+2\left(2 \epsilon^2+\mu^2\right) \cos ^2 \frac{\theta}{2}}{E^2-E_{+}^2 / 4}+\left(\cos \frac{\theta}{2} \rightarrow \sin \frac{\theta}{2}\right)\right)\right]
\end{equation}

\begin{equation}
c_2(\theta)=\frac{\mu}{5 \epsilon+3 \mu}\left[-2 \frac{\epsilon^2+3 \mu \epsilon+\mu^2}{\mu \epsilon}+\frac{2 \epsilon+\mu}{\epsilon}\left(\frac{2 \epsilon+\mu}{2 \epsilon \cos ^2 \frac{\theta}{2}+\mu}+\left(\cos \frac{\theta}{2} \rightarrow \sin \frac{\theta}{2}\right)\right)\right],
\end{equation}
\end{widetext}
where $E_\pm = \sqrt{\epsilon_{\mathbf{k}\pm\mathbf{p}} (\epsilon_{\mathbf{k}\pm\mathbf{p}} + 2\mu)}$ with $\epsilon_{\mathbf{k}+\mathbf{p}} = 4\epsilon\cos^2\frac{\theta}{2}$ and $\epsilon_{\mathbf{k}-\mathbf{p}} = 4\epsilon\sin^2\frac{\theta}{2}$. Performing small perturbations around the lattice fixed point, one can linearize eq.\,(\ref{eq:GinzburgLandau}), and extract growth rates of these small excitations to define a stability criterion \cite{Keisuke}
\begin{equation}
D = -1 + c_1(\theta)^2 + 2 c_2(\theta) - c_2(\theta)^2,
\end{equation}
where $D < 0$ represents a stable fixed point and $D > 0$ an unstable point. The functional form is plotted in \cref{fig:ExtendedData}.
Angles $\sim\,90^\mathrm{o}$ are always stable, though the band becomes narrower for higher drive frequencies. 
Divergences occur at the angle where the energy of outgoing quasi-particles of a collision are resonant to the drive energy, i.e. $E_{k+p} = \omega_d$. 
This divergence is discussed in detail in \cite{Keisuke}, and has no physical reality.

\section*{Appendix E: Extraction of Experimental Parameters}
In the following, we determine the growth rate $\alpha$, damping $\Gamma$, and the effective $\lambda$ in the experiment. The predicted location of the fixed point is given by \cite{Keisuke}
\begin{align}
\label{eq:fixedpoint}
    |R_{\mathrm{fp}}|^2 = \frac{\Gamma}{\lambda}\;\frac{\sqrt{(\frac{\alpha}{\Gamma})^2 - 1}}{1 + c_1 + c_2}.
\end{align}
Factors $\alpha$, $\lambda$, $c_1$, and $c_2$ each depend on the driving frequency, and at higher drive frequencies the fixed point is at larger values of $R$. Only $\alpha$ additionally depends on the drive amplitude $r$. We fit the function $F(r) = a\sqrt{(r/b)^2-1}$ to the data plotted in \cref{fig:Fig3}c. We extract $b = 0.19\pm0.01$, which is the threshold $r$ below which no quasi-particles are produced.  We can conclude that $\Gamma = \alpha(r = 0.19)$. Next, we use the growth rate extracted from \cref{fig:Fig2}b where
\begin{align}
    n(t) = e^{2(\alpha - \Gamma)t}n(t=0),
\end{align}
with shake amplitude $r = 0.4$, and find $2(\alpha - \Gamma) = (95 \pm 10) \mathrm{s}^{-1}$. By using the growth rate and the critical drive amplitude, we find $\Gamma = (44 \pm 5)\mathrm{s}^{-1}$ and $\alpha(r = 0.4) = (90 \pm 10)\mathrm{s}^{-1}$, which is in good agreement with the theory value of 114 $\mathrm{s}^{-1}$.
Because we measure the occupations at the time where \cref{eq:6} is maximal and we integrate over all four momentum peaks, we rewrite \cref{eq:fixedpoint} for momentum space occupations
\begin{align}
   n_{\mathrm{fp}}(k_c) = 4 N_0 \left(1 + 2 \frac{\mu}{\epsilon}\right)\frac{\Gamma\sqrt{(\frac{\alpha}{\Gamma})^2 - 1}}{\lambda(1 + c_1 + c_2)}.
\end{align}
We compare this equation with the fit function $F$ and use fit parameter $a$ to find
\begin{align}
    \lambda_{\mathrm{eff}} = \frac{4 N_0 (1 + 2 \frac{\mu}{\epsilon})\Gamma}{(1 + c_1 + c_2)} \frac{1}{a},
\end{align}
yielding  $\lambda_{\mathrm{eff}} = (3.4 \pm 0.4)\times10^4 \mathrm{s}^{-1}$. The error is dominated by the uncertainty in $\Gamma$, which is given by the uncertainty in the growth rate. This is a factor 3 larger than the theoretical value. We likely underestimate the absolute value, as integration over the resonant momenta includes incoherent particles as well as excitations beyond the two-stripe description. We use $\lambda_{\mathrm{eff}}$ for generation of all flow plots and stability diagrams.
\\
\\
\section*{Appendix F: Amplitude Equation in Polar Form}
 One can rewrite eq.\,(\ref{eq:GinzburgLandau}) such that dynamics of the magnitude and phase of the phonons are separated, 
\begin{eqnarray}
    \label{eq:GinzburgLandauAmpAndPhase}
    \frac{d}{d t}\left|R_k(t)\right|=&& \Big[-\alpha \cos\left(2\varphi_k(t)\right) -\Gamma \\ \nonumber
    &&
    +\lambda c_2(\theta)\left|R_p(t)\right|^2 \sin \left(2 \varphi_p(t)-2 \varphi_k(t)\right)\Big]\left|R_k(t)\right|\\ \nonumber
\end{eqnarray}
\begin{eqnarray}
\frac{d}{d t} \varphi_k(t) =&& \alpha \sin \left(2 \varphi_k(t)\right)-\lambda\Bigg[\left|R_k(t)\right|^2 \\ \nonumber
&&
+\Big(c_1(\theta)+c_2(\theta) \cos \big(2 \varphi_p(t)-2 \varphi_k(t)\big)\Big)\left|R_p(t)\right|^2\Bigg].
\end{eqnarray}
Here, one can separate the dynamics of the amplitudes (\cref{fig:Fig4}c and f) and the phases (\cref{fig:Fig4}c and f insets) of phonons. Here, the mechanism for limiting amplitudes even of single density waves can be understood: large amplitudes induce a phase shift relative to the drive, which in turn causes growth to cease. This induced phase lag to the shaking can be seen clearly in \cref{fig:Fig4}c.

\newpage
\bibliography{references.bib}

\begin{thebibliography}{31}%
\makeatletter
\providecommand \@ifxundefined [1]{%
 \@ifx{#1\undefined}
}%
\providecommand \@ifnum [1]{%
 \ifnum #1\expandafter \@firstoftwo
 \else \expandafter \@secondoftwo
 \fi
}%
\providecommand \@ifx [1]{%
 \ifx #1\expandafter \@firstoftwo
 \else \expandafter \@secondoftwo
 \fi
}%
\providecommand \natexlab [1]{#1}%
\providecommand \enquote  [1]{``#1''}%
\providecommand \bibnamefont  [1]{#1}%
\providecommand \bibfnamefont [1]{#1}%
\providecommand \citenamefont [1]{#1}%
\providecommand \href@noop [0]{\@secondoftwo}%
\providecommand \href [0]{\begingroup \@sanitize@url \@href}%
\providecommand \@href[1]{\@@startlink{#1}\@@href}%
\providecommand \@@href[1]{\endgroup#1\@@endlink}%
\providecommand \@sanitize@url [0]{\catcode `\\12\catcode `\$12\catcode
  `\&12\catcode `\#12\catcode `\^12\catcode `\_12\catcode `\%12\relax}%
\providecommand \@@startlink[1]{}%
\providecommand \@@endlink[0]{}%
\providecommand \url  [0]{\begingroup\@sanitize@url \@url }%
\providecommand \@url [1]{\endgroup\@href {#1}{\urlprefix }}%
\providecommand \urlprefix  [0]{URL }%
\providecommand \Eprint [0]{\href }%
\providecommand \doibase [0]{https://doi.org/}%
\providecommand \selectlanguage [0]{\@gobble}%
\providecommand \bibinfo  [0]{\@secondoftwo}%
\providecommand \bibfield  [0]{\@secondoftwo}%
\providecommand \translation [1]{[#1]}%
\providecommand \BibitemOpen [0]{}%
\providecommand \bibitemStop [0]{}%
\providecommand \bibitemNoStop [0]{.\EOS\space}%
\providecommand \EOS [0]{\spacefactor3000\relax}%
\providecommand \BibitemShut  [1]{\csname bibitem#1\endcsname}%
\let\auto@bib@innerbib\@empty
\bibitem [{\citenamefont {Cross}\ and\ \citenamefont
  {Hohenberg}(1993)}]{CrossHohenberg}%
  \BibitemOpen
  \bibfield  {author} {\bibinfo {author} {\bibfnamefont {M.~C.}\ \bibnamefont
  {Cross}}\ and\ \bibinfo {author} {\bibfnamefont {P.~C.}\ \bibnamefont
  {Hohenberg}},\ }\bibfield  {title} {\bibinfo {title} {Pattern formation
  outside of equilibrium},\ }\href {https://doi.org/10.1103/RevModPhys.65.851}
  {\bibfield  {journal} {\bibinfo  {journal} {Rev. Mod. Phys.}\ }\textbf
  {\bibinfo {volume} {65}},\ \bibinfo {pages} {851} (\bibinfo {year}
  {1993})}\BibitemShut {NoStop}%
\bibitem [{\citenamefont {Arecchi}\ \emph {et~al.}(1999)\citenamefont
  {Arecchi}, \citenamefont {Boccaletti},\ and\ \citenamefont
  {Ramazza}}]{ARECCHI19991}%
  \BibitemOpen
  \bibfield  {author} {\bibinfo {author} {\bibfnamefont {F.}~\bibnamefont
  {Arecchi}}, \bibinfo {author} {\bibfnamefont {S.}~\bibnamefont
  {Boccaletti}},\ and\ \bibinfo {author} {\bibfnamefont {P.}~\bibnamefont
  {Ramazza}},\ }\bibfield  {title} {\bibinfo {title} {Pattern formation and
  competition in nonlinear optics},\ }\href
  {https://doi.org/https://doi.org/10.1016/S0370-1573(99)00007-1} {\bibfield
  {journal} {\bibinfo  {journal} {Physics Reports}\ }\textbf {\bibinfo {volume}
  {318}},\ \bibinfo {pages} {1} (\bibinfo {year} {1999})}\BibitemShut {NoStop}%
\bibitem [{\citenamefont {Koch}\ and\ \citenamefont
  {Meinhardt}(1994)}]{Koch1994}%
  \BibitemOpen
  \bibfield  {author} {\bibinfo {author} {\bibfnamefont {A.~J.}\ \bibnamefont
  {Koch}}\ and\ \bibinfo {author} {\bibfnamefont {H.}~\bibnamefont
  {Meinhardt}},\ }\bibfield  {title} {\bibinfo {title} {Biological pattern
  formation: from basic mechanisms to complex structures},\ }\href
  {https://doi.org/10.1103/RevModPhys.66.1481} {\bibfield  {journal} {\bibinfo
  {journal} {Rev. Mod. Phys.}\ }\textbf {\bibinfo {volume} {66}},\ \bibinfo
  {pages} {1481} (\bibinfo {year} {1994})}\BibitemShut {NoStop}%
\bibitem [{\citenamefont {K.~Maini}\ \emph {et~al.}(1997)\citenamefont
  {K.~Maini}, \citenamefont {J.~Painter},\ and\ \citenamefont {Nguyen
  Phong~Chau}}]{Maini1997}%
  \BibitemOpen
  \bibfield  {author} {\bibinfo {author} {\bibfnamefont {P.}~\bibnamefont
  {K.~Maini}}, \bibinfo {author} {\bibfnamefont {K.}~\bibnamefont
  {J.~Painter}},\ and\ \bibinfo {author} {\bibfnamefont {H.}~\bibnamefont
  {Nguyen Phong~Chau}},\ }\bibfield  {title} {\bibinfo {title} {Spatial pattern
  formation in chemical and biological systems},\ }\href
  {https://doi.org/10.1039/A702602A} {\bibfield  {journal} {\bibinfo  {journal}
  {J. Chem. Soc.{,} Faraday Trans.}\ }\textbf {\bibinfo {volume} {93}},\
  \bibinfo {pages} {3601} (\bibinfo {year} {1997})}\BibitemShut {NoStop}%
\bibitem [{\citenamefont {Cross}\ and\ \citenamefont
  {Greenside}(2009)}]{cross2009pattern}%
  \BibitemOpen
  \bibfield  {author} {\bibinfo {author} {\bibfnamefont {M.}~\bibnamefont
  {Cross}}\ and\ \bibinfo {author} {\bibfnamefont {H.}~\bibnamefont
  {Greenside}},\ }\href {https://books.google.de/books?id=s04bhzQfPcwC} {\emph
  {\bibinfo {title} {Pattern Formation and Dynamics in Nonequilibrium
  Systems}}}\ (\bibinfo  {publisher} {Cambridge University Press},\ \bibinfo
  {year} {2009})\BibitemShut {NoStop}%
\bibitem [{\citenamefont {Milner}(1991)}]{Milner_1991}%
  \BibitemOpen
  \bibfield  {author} {\bibinfo {author} {\bibfnamefont {S.~T.}\ \bibnamefont
  {Milner}},\ }\bibfield  {title} {\bibinfo {title} {Square patterns and
  secondary instabilities in driven capillary waves},\ }\href
  {https://doi.org/10.1017/S0022112091001970} {\bibfield  {journal} {\bibinfo
  {journal} {Journal of Fluid Mechanics}\ }\textbf {\bibinfo {volume} {225}},\
  \bibinfo {pages} {81–100} (\bibinfo {year} {1991})}\BibitemShut {NoStop}%
\bibitem [{\citenamefont {Zhang}\ and\ \citenamefont
  {Viñals}(1997)}]{ZHANG_1997}%
  \BibitemOpen
  \bibfield  {author} {\bibinfo {author} {\bibfnamefont {W.}~\bibnamefont
  {Zhang}}\ and\ \bibinfo {author} {\bibfnamefont {J.}~\bibnamefont
  {Viñals}},\ }\bibfield  {title} {\bibinfo {title} {Pattern formation in
  weakly damped parametric surface waves},\ }\href
  {https://doi.org/10.1017/S0022112096004764} {\bibfield  {journal} {\bibinfo
  {journal} {Journal of Fluid Mechanics}\ }\textbf {\bibinfo {volume} {336}},\
  \bibinfo {pages} {301–330} (\bibinfo {year} {1997})}\BibitemShut {NoStop}%
\bibitem [{\citenamefont {Kevrekidis}\ and\ \citenamefont
  {Frantzeskakis}(2004)}]{Kevrekidis:2004}%
  \BibitemOpen
  \bibfield  {author} {\bibinfo {author} {\bibfnamefont {P.~G.}\ \bibnamefont
  {Kevrekidis}}\ and\ \bibinfo {author} {\bibfnamefont {D.~J.}\ \bibnamefont
  {Frantzeskakis}},\ }\bibfield  {title} {\bibinfo {title} {Pattern forming
  dynamical instabilities of bose-einstein condensates},\ }\href
  {https://doi.org/10.1142/S0217984904006809} {\bibfield  {journal} {\bibinfo
  {journal} {Mod. Phys. Lett. B}\ }\textbf {\bibinfo {volume} {18}},\ \bibinfo
  {pages} {173} (\bibinfo {year} {2004})}\BibitemShut {NoStop}%
\bibitem [{\citenamefont {Engels}\ \emph {et~al.}(2007)\citenamefont {Engels},
  \citenamefont {Atherton},\ and\ \citenamefont
  {Hoefer}}]{Engels2007_FaradayinBEC}%
  \BibitemOpen
  \bibfield  {author} {\bibinfo {author} {\bibfnamefont {P.}~\bibnamefont
  {Engels}}, \bibinfo {author} {\bibfnamefont {C.}~\bibnamefont {Atherton}},\
  and\ \bibinfo {author} {\bibfnamefont {M.~A.}\ \bibnamefont {Hoefer}},\
  }\bibfield  {title} {\bibinfo {title} {Observation of faraday waves in a
  bose-einstein condensate},\ }\href
  {https://doi.org/10.1103/PhysRevLett.98.095301} {\bibfield  {journal}
  {\bibinfo  {journal} {Phys. Rev. Lett.}\ }\textbf {\bibinfo {volume} {98}},\
  \bibinfo {pages} {095301} (\bibinfo {year} {2007})}\BibitemShut {NoStop}%
\bibitem [{\citenamefont {Nicolin}\ \emph {et~al.}(2007)\citenamefont
  {Nicolin}, \citenamefont {Carretero-Gonz\'alez},\ and\ \citenamefont
  {Kevrekidis}}]{Nicolin:2007}%
  \BibitemOpen
  \bibfield  {author} {\bibinfo {author} {\bibfnamefont {A.~I.}\ \bibnamefont
  {Nicolin}}, \bibinfo {author} {\bibfnamefont {R.}~\bibnamefont
  {Carretero-Gonz\'alez}},\ and\ \bibinfo {author} {\bibfnamefont {P.~G.}\
  \bibnamefont {Kevrekidis}},\ }\bibfield  {title} {\bibinfo {title} {Faraday
  waves in bose-einstein condensates},\ }\href
  {https://doi.org/10.1103/PhysRevA.76.063609} {\bibfield  {journal} {\bibinfo
  {journal} {Phys. Rev. A}\ }\textbf {\bibinfo {volume} {76}},\ \bibinfo
  {pages} {063609} (\bibinfo {year} {2007})}\BibitemShut {NoStop}%
\bibitem [{\citenamefont {Capuzzi}\ and\ \citenamefont
  {Vignolo}(2008)}]{Capuzzi:2008}%
  \BibitemOpen
  \bibfield  {author} {\bibinfo {author} {\bibfnamefont {P.}~\bibnamefont
  {Capuzzi}}\ and\ \bibinfo {author} {\bibfnamefont {P.}~\bibnamefont
  {Vignolo}},\ }\bibfield  {title} {\bibinfo {title} {Faraday waves in
  elongated superfluid fermionic clouds},\ }\href
  {https://doi.org/10.1103/PhysRevA.78.043613} {\bibfield  {journal} {\bibinfo
  {journal} {Phys. Rev. A}\ }\textbf {\bibinfo {volume} {78}},\ \bibinfo
  {pages} {043613} (\bibinfo {year} {2008})}\BibitemShut {NoStop}%
\bibitem [{\citenamefont {Tang}\ \emph {et~al.}(2011)\citenamefont {Tang},
  \citenamefont {Li},\ and\ \citenamefont {Xue}}]{Tang:2011}%
  \BibitemOpen
  \bibfield  {author} {\bibinfo {author} {\bibfnamefont {R.-A.}\ \bibnamefont
  {Tang}}, \bibinfo {author} {\bibfnamefont {H.-C.}\ \bibnamefont {Li}},\ and\
  \bibinfo {author} {\bibfnamefont {J.-K.}\ \bibnamefont {Xue}},\ }\bibfield
  {title} {\bibinfo {title} {Faraday instability and faraday patterns in a
  superfluid fermi gas},\ }\href
  {https://dx.doi.org/10.1088/0953-4075/44/11/115303} {\bibfield  {journal}
  {\bibinfo  {journal} {J. Phys. B: At. Mol. Opt. Phys.}\ }\textbf {\bibinfo
  {volume} {44}},\ \bibinfo {pages} {115303} (\bibinfo {year}
  {2011})}\BibitemShut {NoStop}%
\bibitem [{\citenamefont {Nicolin}(lack)}]{Nicolin2011}%
  \BibitemOpen
  \bibfield  {author} {\bibinfo {author} {\bibfnamefont {A.~I.}\ \bibnamefont
  {Nicolin}},\ }\bibfield  {title} {\bibinfo {title} {Resonant wave formation
  in bose-einstein condensates},\ }\href
  {https://doi.org/10.1103/PhysRevE.84.056202} {\bibfield  {journal} {\bibinfo
  {journal} {Phys. Rev. E}\ }\textbf {\bibinfo {volume} {84}},\ \bibinfo
  {pages} {056202} (\bibinfo {year} {2011\color{black}})}\BibitemShut {NoStop}%
\bibitem [{\citenamefont {Jaskula}\ \emph {et~al.}(2012)\citenamefont
  {Jaskula}, \citenamefont {Partridge}, \citenamefont {Bonneau}, \citenamefont
  {Lopes}, \citenamefont {Ruaudel}, \citenamefont {Boiron},\ and\ \citenamefont
  {Westbrook}}]{Westbrook2012_DynaCasi}%
  \BibitemOpen
  \bibfield  {author} {\bibinfo {author} {\bibfnamefont {J.-C.}\ \bibnamefont
  {Jaskula}}, \bibinfo {author} {\bibfnamefont {G.~B.}\ \bibnamefont
  {Partridge}}, \bibinfo {author} {\bibfnamefont {M.}~\bibnamefont {Bonneau}},
  \bibinfo {author} {\bibfnamefont {R.}~\bibnamefont {Lopes}}, \bibinfo
  {author} {\bibfnamefont {J.}~\bibnamefont {Ruaudel}}, \bibinfo {author}
  {\bibfnamefont {D.}~\bibnamefont {Boiron}},\ and\ \bibinfo {author}
  {\bibfnamefont {C.~I.}\ \bibnamefont {Westbrook}},\ }\bibfield  {title}
  {\bibinfo {title} {Acoustic analog to the dynamical casimir effect in a
  bose-einstein condensate},\ }\href
  {https://doi.org/10.1103/PhysRevLett.109.220401} {\bibfield  {journal}
  {\bibinfo  {journal} {Phys. Rev. Lett.}\ }\textbf {\bibinfo {volume} {109}},\
  \bibinfo {pages} {220401} (\bibinfo {year} {2012})}\BibitemShut {NoStop}%
\bibitem [{\citenamefont {Smits}\ \emph {et~al.}(2018)\citenamefont {Smits},
  \citenamefont {Liao}, \citenamefont {Stoof},\ and\ \citenamefont {van~der
  Straten}}]{Smits2018_SpaceTimeCrystal}%
  \BibitemOpen
  \bibfield  {author} {\bibinfo {author} {\bibfnamefont {J.}~\bibnamefont
  {Smits}}, \bibinfo {author} {\bibfnamefont {L.}~\bibnamefont {Liao}},
  \bibinfo {author} {\bibfnamefont {H.~T.~C.}\ \bibnamefont {Stoof}},\ and\
  \bibinfo {author} {\bibfnamefont {P.}~\bibnamefont {van~der Straten}},\
  }\bibfield  {title} {\bibinfo {title} {Observation of a space-time crystal in
  a superfluid quantum gas},\ }\href
  {https://doi.org/10.1103/PhysRevLett.121.185301} {\bibfield  {journal}
  {\bibinfo  {journal} {Phys. Rev. Lett.}\ }\textbf {\bibinfo {volume} {121}},\
  \bibinfo {pages} {185301} (\bibinfo {year} {2018})}\BibitemShut {NoStop}%
\bibitem [{\citenamefont {Nguyen}\ \emph {et~al.}(2019)\citenamefont {Nguyen},
  \citenamefont {Tsatsos}, \citenamefont {Luo}, \citenamefont {Lode},
  \citenamefont {Telles}, \citenamefont {Bagnato},\ and\ \citenamefont
  {Hulet}}]{Jason2019}%
  \BibitemOpen
  \bibfield  {author} {\bibinfo {author} {\bibfnamefont {J.~H.~V.}\
  \bibnamefont {Nguyen}}, \bibinfo {author} {\bibfnamefont {M.~C.}\
  \bibnamefont {Tsatsos}}, \bibinfo {author} {\bibfnamefont {D.}~\bibnamefont
  {Luo}}, \bibinfo {author} {\bibfnamefont {A.~U.~J.}\ \bibnamefont {Lode}},
  \bibinfo {author} {\bibfnamefont {G.~D.}\ \bibnamefont {Telles}}, \bibinfo
  {author} {\bibfnamefont {V.~S.}\ \bibnamefont {Bagnato}},\ and\ \bibinfo
  {author} {\bibfnamefont {R.~G.}\ \bibnamefont {Hulet}},\ }\bibfield  {title}
  {\bibinfo {title} {Parametric excitation of a bose-einstein condensate: From
  faraday waves to granulation},\ }\href
  {https://doi.org/10.1103/PhysRevX.9.011052} {\bibfield  {journal} {\bibinfo
  {journal} {Phys. Rev. X}\ }\textbf {\bibinfo {volume} {9}},\ \bibinfo {pages}
  {011052} (\bibinfo {year} {2019})}\BibitemShut {NoStop}%
\bibitem [{\citenamefont {Hernández-Rajkov}\ \emph {et~al.}(2021)\citenamefont
  {Hernández-Rajkov}, \citenamefont {Padilla-Castillo}, \citenamefont {del
  Río-Lima}, \citenamefont {Gutiérrez-Valdés}, \citenamefont
  {Poveda-Cuevas},\ and\ \citenamefont {Seman}}]{HernandezRajkov_2021}%
  \BibitemOpen
  \bibfield  {author} {\bibinfo {author} {\bibfnamefont {D.}~\bibnamefont
  {Hernández-Rajkov}}, \bibinfo {author} {\bibfnamefont {J.~E.}\ \bibnamefont
  {Padilla-Castillo}}, \bibinfo {author} {\bibfnamefont {A.}~\bibnamefont {del
  Río-Lima}}, \bibinfo {author} {\bibfnamefont {A.}~\bibnamefont
  {Gutiérrez-Valdés}}, \bibinfo {author} {\bibfnamefont {F.~J.}\ \bibnamefont
  {Poveda-Cuevas}},\ and\ \bibinfo {author} {\bibfnamefont {J.~A.}\
  \bibnamefont {Seman}},\ }\bibfield  {title} {\bibinfo {title} {Faraday waves
  in strongly interacting superfluids},\ }\href
  {https://doi.org/10.1088/1367-2630/ac2d70} {\bibfield  {journal} {\bibinfo
  {journal} {New Journal of Physics}\ }\textbf {\bibinfo {volume} {23}},\
  \bibinfo {pages} {103038} (\bibinfo {year} {2021})}\BibitemShut {NoStop}%
\bibitem [{\citenamefont {Dupont}\ \emph {et~al.}(2023)\citenamefont {Dupont},
  \citenamefont {Gabardos}, \citenamefont {Arrouas}, \citenamefont {Chatelain},
  \citenamefont {Arnal}, \citenamefont {Billy}, \citenamefont {Schlagheck},
  \citenamefont {Peaudecerf},\ and\ \citenamefont
  {Guéry-Odelin}}]{Dupont2023}%
  \BibitemOpen
  \bibfield  {author} {\bibinfo {author} {\bibfnamefont {N.}~\bibnamefont
  {Dupont}}, \bibinfo {author} {\bibfnamefont {L.}~\bibnamefont {Gabardos}},
  \bibinfo {author} {\bibfnamefont {F.}~\bibnamefont {Arrouas}}, \bibinfo
  {author} {\bibfnamefont {G.}~\bibnamefont {Chatelain}}, \bibinfo {author}
  {\bibfnamefont {M.}~\bibnamefont {Arnal}}, \bibinfo {author} {\bibfnamefont
  {J.}~\bibnamefont {Billy}}, \bibinfo {author} {\bibfnamefont
  {P.}~\bibnamefont {Schlagheck}}, \bibinfo {author} {\bibfnamefont
  {B.}~\bibnamefont {Peaudecerf}},\ and\ \bibinfo {author} {\bibfnamefont
  {D.}~\bibnamefont {Guéry-Odelin}},\ }\bibfield  {title} {\bibinfo {title}
  {Emergence of tunable periodic density correlations in a floquet–bloch
  system},\ }\href {https://doi.org/10.1073/pnas.2300980120} {\bibfield
  {journal} {\bibinfo  {journal} {Proceedings of the National Academy of
  Sciences}\ }\textbf {\bibinfo {volume} {120}},\ \bibinfo {pages}
  {e2300980120} (\bibinfo {year} {2023})},\ \Eprint
  {https://arxiv.org/abs/https://www.pnas.org/doi/pdf/10.1073/pnas.2300980120}
  {https://www.pnas.org/doi/pdf/10.1073/pnas.2300980120} \BibitemShut {NoStop}%
\bibitem [{\citenamefont {Smits}\ \emph {et~al.}(2020)\citenamefont {Smits},
  \citenamefont {Stoof},\ and\ \citenamefont {van~der Straten}}]{Smits_2020}%
  \BibitemOpen
  \bibfield  {author} {\bibinfo {author} {\bibfnamefont {J.}~\bibnamefont
  {Smits}}, \bibinfo {author} {\bibfnamefont {H.~T.~C.}\ \bibnamefont
  {Stoof}},\ and\ \bibinfo {author} {\bibfnamefont {P.}~\bibnamefont {van~der
  Straten}},\ }\bibfield  {title} {\bibinfo {title} {On the long-term stability
  of space-time crystals},\ }\href {https://doi.org/10.1088/1367-2630/abbae9}
  {\bibfield  {journal} {\bibinfo  {journal} {New Journal of Physics}\ }\textbf
  {\bibinfo {volume} {22}},\ \bibinfo {pages} {105001} (\bibinfo {year}
  {2020})}\BibitemShut {NoStop}%
\bibitem [{\citenamefont {Zhang}\ \emph {et~al.}(2020)\citenamefont {Zhang},
  \citenamefont {Yao}, \citenamefont {Feng}, \citenamefont {Hu},\ and\
  \citenamefont {Chin}}]{Zhang2020}%
  \BibitemOpen
  \bibfield  {author} {\bibinfo {author} {\bibfnamefont {Z.}~\bibnamefont
  {Zhang}}, \bibinfo {author} {\bibfnamefont {K.~X.}\ \bibnamefont {Yao}},
  \bibinfo {author} {\bibfnamefont {L.}~\bibnamefont {Feng}}, \bibinfo {author}
  {\bibfnamefont {J.}~\bibnamefont {Hu}},\ and\ \bibinfo {author}
  {\bibfnamefont {C.}~\bibnamefont {Chin}},\ }\bibfield  {title} {\bibinfo
  {title} {Pattern formation in a driven bose–einstein condensate},\ }\href
  {https://doi.org/10.1038/s41567-020-0839-3} {\bibfield  {journal} {\bibinfo
  {journal} {Nature Physics}\ }\textbf {\bibinfo {volume} {16}},\ \bibinfo
  {pages} {652} (\bibinfo {year} {2020})}\BibitemShut {NoStop}%
\bibitem [{\citenamefont {Kwon}\ \emph {et~al.}(2021)\citenamefont {Kwon},
  \citenamefont {Mukherjee}, \citenamefont {Huh}, \citenamefont {Kim},
  \citenamefont {Mistakidis}, \citenamefont {Maity}, \citenamefont
  {Kevrekidis}, \citenamefont {Majumder}, \citenamefont {Schmelcher},\ and\
  \citenamefont {Choi}}]{Kwon2021}%
  \BibitemOpen
  \bibfield  {author} {\bibinfo {author} {\bibfnamefont {K.}~\bibnamefont
  {Kwon}}, \bibinfo {author} {\bibfnamefont {K.}~\bibnamefont {Mukherjee}},
  \bibinfo {author} {\bibfnamefont {S.~J.}\ \bibnamefont {Huh}}, \bibinfo
  {author} {\bibfnamefont {K.}~\bibnamefont {Kim}}, \bibinfo {author}
  {\bibfnamefont {S.~I.}\ \bibnamefont {Mistakidis}}, \bibinfo {author}
  {\bibfnamefont {D.~K.}\ \bibnamefont {Maity}}, \bibinfo {author}
  {\bibfnamefont {P.~G.}\ \bibnamefont {Kevrekidis}}, \bibinfo {author}
  {\bibfnamefont {S.}~\bibnamefont {Majumder}}, \bibinfo {author}
  {\bibfnamefont {P.}~\bibnamefont {Schmelcher}},\ and\ \bibinfo {author}
  {\bibfnamefont {J.-y.}\ \bibnamefont {Choi}},\ }\bibfield  {title} {\bibinfo
  {title} {Spontaneous formation of star-shaped surface patterns in a driven
  bose-einstein condensate},\ }\href
  {https://doi.org/10.1103/PhysRevLett.127.113001} {\bibfield  {journal}
  {\bibinfo  {journal} {Phys. Rev. Lett.}\ }\textbf {\bibinfo {volume} {127}},\
  \bibinfo {pages} {113001} (\bibinfo {year} {2021})}\BibitemShut {NoStop}%
\bibitem [{\citenamefont {Etrych}\ \emph {et~al.}(2023)\citenamefont {Etrych},
  \citenamefont {Martirosyan}, \citenamefont {Cao}, \citenamefont {Glidden},
  \citenamefont {Dogra}, \citenamefont {Hutson}, \citenamefont {Hadzibabic},\
  and\ \citenamefont {Eigen}}]{HadziFeshbach}%
  \BibitemOpen
  \bibfield  {author} {\bibinfo {author} {\bibfnamefont {J.~c.~v.}\
  \bibnamefont {Etrych}}, \bibinfo {author} {\bibfnamefont {G.}~\bibnamefont
  {Martirosyan}}, \bibinfo {author} {\bibfnamefont {A.}~\bibnamefont {Cao}},
  \bibinfo {author} {\bibfnamefont {J.~A.~P.}\ \bibnamefont {Glidden}},
  \bibinfo {author} {\bibfnamefont {L.~H.}\ \bibnamefont {Dogra}}, \bibinfo
  {author} {\bibfnamefont {J.~M.}\ \bibnamefont {Hutson}}, \bibinfo {author}
  {\bibfnamefont {Z.}~\bibnamefont {Hadzibabic}},\ and\ \bibinfo {author}
  {\bibfnamefont {C.}~\bibnamefont {Eigen}},\ }\bibfield  {title} {\bibinfo
  {title} {Pinpointing feshbach resonances and testing efimov universalities in
  $^{39}\mathrm{K}$},\ }\href
  {https://doi.org/10.1103/PhysRevResearch.5.013174} {\bibfield  {journal}
  {\bibinfo  {journal} {Phys. Rev. Res.}\ }\textbf {\bibinfo {volume} {5}},\
  \bibinfo {pages} {013174} (\bibinfo {year} {2023})}\BibitemShut {NoStop}%
\bibitem [{\citenamefont {Hans}\ \emph {et~al.}(2021)\citenamefont {Hans},
  \citenamefont {Schmutte}, \citenamefont {Viermann}, \citenamefont {Liebster},
  \citenamefont {Sparn}, \citenamefont {Oberthaler},\ and\ \citenamefont
  {Strobel}}]{Hans2021}%
  \BibitemOpen
  \bibfield  {author} {\bibinfo {author} {\bibfnamefont {M.}~\bibnamefont
  {Hans}}, \bibinfo {author} {\bibfnamefont {F.}~\bibnamefont {Schmutte}},
  \bibinfo {author} {\bibfnamefont {C.}~\bibnamefont {Viermann}}, \bibinfo
  {author} {\bibfnamefont {N.}~\bibnamefont {Liebster}}, \bibinfo {author}
  {\bibfnamefont {M.}~\bibnamefont {Sparn}}, \bibinfo {author} {\bibfnamefont
  {M.~K.}\ \bibnamefont {Oberthaler}},\ and\ \bibinfo {author} {\bibfnamefont
  {H.}~\bibnamefont {Strobel}},\ }\bibfield  {title} {\bibinfo {title} {High
  signal to noise absorption imaging of alkali atoms at moderate magnetic
  fields},\ }\href {https://doi.org/10.1063/5.0040677} {\bibfield  {journal}
  {\bibinfo  {journal} {Review of Scientific Instruments}\ }\textbf {\bibinfo
  {volume} {92}},\ \bibinfo {pages} {023203} (\bibinfo {year} {2021})},\
  \Eprint {https://arxiv.org/abs/https://doi.org/10.1063/5.0040677}
  {https://doi.org/10.1063/5.0040677} \BibitemShut {NoStop}%
\bibitem [{\citenamefont {Tung}\ \emph {et~al.}(2010)\citenamefont {Tung},
  \citenamefont {Lamporesi}, \citenamefont {Lobser}, \citenamefont {Xia},\ and\
  \citenamefont {Cornell}}]{MomentumSpaceCornell}%
  \BibitemOpen
  \bibfield  {author} {\bibinfo {author} {\bibfnamefont {S.}~\bibnamefont
  {Tung}}, \bibinfo {author} {\bibfnamefont {G.}~\bibnamefont {Lamporesi}},
  \bibinfo {author} {\bibfnamefont {D.}~\bibnamefont {Lobser}}, \bibinfo
  {author} {\bibfnamefont {L.}~\bibnamefont {Xia}},\ and\ \bibinfo {author}
  {\bibfnamefont {E.~A.}\ \bibnamefont {Cornell}},\ }\bibfield  {title}
  {\bibinfo {title} {Observation of the presuperfluid regime in a
  two-dimensional bose gas},\ }\href
  {https://doi.org/10.1103/PhysRevLett.105.230408} {\bibfield  {journal}
  {\bibinfo  {journal} {Phys. Rev. Lett.}\ }\textbf {\bibinfo {volume} {105}},\
  \bibinfo {pages} {230408} (\bibinfo {year} {2010})}\BibitemShut {NoStop}%
\bibitem [{\citenamefont {Murthy}\ \emph {et~al.}(2014)\citenamefont {Murthy},
  \citenamefont {Kedar}, \citenamefont {Lompe}, \citenamefont {Neidig},
  \citenamefont {Ries}, \citenamefont {Wenz}, \citenamefont {Z\"urn},\ and\
  \citenamefont {Jochim}}]{Murthy2014}%
  \BibitemOpen
  \bibfield  {author} {\bibinfo {author} {\bibfnamefont {P.~A.}\ \bibnamefont
  {Murthy}}, \bibinfo {author} {\bibfnamefont {D.}~\bibnamefont {Kedar}},
  \bibinfo {author} {\bibfnamefont {T.}~\bibnamefont {Lompe}}, \bibinfo
  {author} {\bibfnamefont {M.}~\bibnamefont {Neidig}}, \bibinfo {author}
  {\bibfnamefont {M.~G.}\ \bibnamefont {Ries}}, \bibinfo {author}
  {\bibfnamefont {A.~N.}\ \bibnamefont {Wenz}}, \bibinfo {author}
  {\bibfnamefont {G.}~\bibnamefont {Z\"urn}},\ and\ \bibinfo {author}
  {\bibfnamefont {S.}~\bibnamefont {Jochim}},\ }\bibfield  {title} {\bibinfo
  {title} {Matter-wave fourier optics with a strongly interacting
  two-dimensional fermi gas},\ }\href
  {https://doi.org/10.1103/PhysRevA.90.043611} {\bibfield  {journal} {\bibinfo
  {journal} {Phys. Rev. A}\ }\textbf {\bibinfo {volume} {90}},\ \bibinfo
  {pages} {043611} (\bibinfo {year} {2014})}\BibitemShut {NoStop}%
\bibitem [{\citenamefont {Faraday}(1837)}]{Faraday:1837}%
  \BibitemOpen
  \bibfield  {author} {\bibinfo {author} {\bibfnamefont {M.}~\bibnamefont
  {Faraday}},\ }\bibfield  {title} {\bibinfo {title} {On a peculiar class of
  acoustical figures; and on certain forms assumed by groups of particles upon
  vibrating elastic surfaces},\ }\href {https://doi.org/10.1098/rspl.1830.0024}
  {\bibfield  {journal} {\bibinfo  {journal} {Phil. Trans. R. Soc. Lond.}\
  }\textbf {\bibinfo {volume} {3}},\ \bibinfo {pages} {49} (\bibinfo {year}
  {1837})}\BibitemShut {NoStop}%
\bibitem [{\citenamefont {Staliunas}\ \emph {et~al.}(2002)\citenamefont
  {Staliunas}, \citenamefont {Longhi},\ and\ \citenamefont
  {de~Valc\'arcel}}]{Staliunas2002}%
  \BibitemOpen
  \bibfield  {author} {\bibinfo {author} {\bibfnamefont {K.}~\bibnamefont
  {Staliunas}}, \bibinfo {author} {\bibfnamefont {S.}~\bibnamefont {Longhi}},\
  and\ \bibinfo {author} {\bibfnamefont {G.~J.}\ \bibnamefont
  {de~Valc\'arcel}},\ }\bibfield  {title} {\bibinfo {title} {Faraday patterns
  in bose-einstein condensates},\ }\href
  {https://doi.org/10.1103/PhysRevLett.89.210406} {\bibfield  {journal}
  {\bibinfo  {journal} {Phys. Rev. Lett.}\ }\textbf {\bibinfo {volume} {89}},\
  \bibinfo {pages} {210406} (\bibinfo {year} {2002})}\BibitemShut {NoStop}%
\bibitem [{\citenamefont {Carusotto}\ \emph {et~al.}(2010)\citenamefont
  {Carusotto}, \citenamefont {Balbinot}, \citenamefont {Fabbri},\ and\
  \citenamefont {Recati}}]{Carusotto2010}%
  \BibitemOpen
  \bibfield  {author} {\bibinfo {author} {\bibfnamefont {I.}~\bibnamefont
  {Carusotto}}, \bibinfo {author} {\bibfnamefont {R.}~\bibnamefont {Balbinot}},
  \bibinfo {author} {\bibfnamefont {A.}~\bibnamefont {Fabbri}},\ and\ \bibinfo
  {author} {\bibfnamefont {A.}~\bibnamefont {Recati}},\ }\bibfield  {title}
  {\bibinfo {title} {Density correlations and analog dynamical casimir emission
  of bogoliubov phonons in modulated atomic bose-einstein condensates},\ }\href
  {https://doi.org/10.1140/epjd/e2009-00314-3} {\bibfield  {journal} {\bibinfo
  {journal} {The European Physical Journal D}\ }\textbf {\bibinfo {volume}
  {56}},\ \bibinfo {pages} {391} (\bibinfo {year} {2010})}\BibitemShut
  {NoStop}%
\bibitem [{\citenamefont {Busch}\ \emph {et~al.}(2014)\citenamefont {Busch},
  \citenamefont {Parentani},\ and\ \citenamefont {Robertson}}]{Busch2014}%
  \BibitemOpen
  \bibfield  {author} {\bibinfo {author} {\bibfnamefont {X.}~\bibnamefont
  {Busch}}, \bibinfo {author} {\bibfnamefont {R.}~\bibnamefont {Parentani}},\
  and\ \bibinfo {author} {\bibfnamefont {S.}~\bibnamefont {Robertson}},\
  }\bibfield  {title} {\bibinfo {title} {Quantum entanglement due to a
  modulated dynamical casimir effect},\ }\href
  {https://doi.org/10.1103/PhysRevA.89.063606} {\bibfield  {journal} {\bibinfo
  {journal} {Phys. Rev. A}\ }\textbf {\bibinfo {volume} {89}},\ \bibinfo
  {pages} {063606} (\bibinfo {year} {2014})}\BibitemShut {NoStop}%
\bibitem [{\citenamefont {Fujii}\ \emph {et~al.}(2024)\citenamefont {Fujii},
  \citenamefont {G\"orlitz}, \citenamefont {Liebster}, \citenamefont {Sparn},
  \citenamefont {Kath}, \citenamefont {Strobel}, \citenamefont {Oberthaler},\
  and\ \citenamefont {Enss}}]{Keisuke}%
  \BibitemOpen
  \bibfield  {author} {\bibinfo {author} {\bibfnamefont {K.}~\bibnamefont
  {Fujii}}, \bibinfo {author} {\bibfnamefont {S.~L.}\ \bibnamefont
  {G\"orlitz}}, \bibinfo {author} {\bibfnamefont {N.}~\bibnamefont {Liebster}},
  \bibinfo {author} {\bibfnamefont {M.}~\bibnamefont {Sparn}}, \bibinfo
  {author} {\bibfnamefont {E.}~\bibnamefont {Kath}}, \bibinfo {author}
  {\bibfnamefont {H.}~\bibnamefont {Strobel}}, \bibinfo {author} {\bibfnamefont
  {M.~K.}\ \bibnamefont {Oberthaler}},\ and\ \bibinfo {author} {\bibfnamefont
  {T.}~\bibnamefont {Enss}},\ }\bibfield  {title} {\bibinfo {title}
  {Stable-fixed-point description of square-pattern formation in driven
  two-dimensional bose-einstein condensates},\ }\href
  {https://doi.org/10.1103/PhysRevA.109.L051301} {\bibfield  {journal}
  {\bibinfo  {journal} {Phys. Rev. A}\ }\textbf {\bibinfo {volume} {109}},\
  \bibinfo {pages} {L051301} (\bibinfo {year} {2024})}\BibitemShut {NoStop}%
\bibitem [{\citenamefont {Viermann}\ \emph {et~al.}(2022)\citenamefont
  {Viermann}, \citenamefont {Sparn}, \citenamefont {Liebster}, \citenamefont
  {Hans}, \citenamefont {Kath}, \citenamefont {Álvaro Parra-López},
  \citenamefont {Tolosa-Simeón}, \citenamefont {Sánchez-Kuntz}, \citenamefont
  {Haas}, \citenamefont {Strobel}, \citenamefont {Floerchinger},\ and\
  \citenamefont {Oberthaler}}]{Viermann2022}%
  \BibitemOpen
  \bibfield  {author} {\bibinfo {author} {\bibfnamefont {C.}~\bibnamefont
  {Viermann}}, \bibinfo {author} {\bibfnamefont {M.}~\bibnamefont {Sparn}},
  \bibinfo {author} {\bibfnamefont {N.}~\bibnamefont {Liebster}}, \bibinfo
  {author} {\bibfnamefont {M.}~\bibnamefont {Hans}}, \bibinfo {author}
  {\bibfnamefont {E.}~\bibnamefont {Kath}}, \bibinfo {author} {\bibnamefont
  {Álvaro Parra-López}}, \bibinfo {author} {\bibfnamefont {M.}~\bibnamefont
  {Tolosa-Simeón}}, \bibinfo {author} {\bibfnamefont {N.}~\bibnamefont
  {Sánchez-Kuntz}}, \bibinfo {author} {\bibfnamefont {T.}~\bibnamefont
  {Haas}}, \bibinfo {author} {\bibfnamefont {H.}~\bibnamefont {Strobel}},
  \bibinfo {author} {\bibfnamefont {S.}~\bibnamefont {Floerchinger}},\ and\
  \bibinfo {author} {\bibfnamefont {M.~K.}\ \bibnamefont {Oberthaler}},\
  }\bibfield  {title} {\bibinfo {title} {Quantum field simulator for dynamics
  in curved spacetime},\ }\href {https://doi.org/10.1038/s41586-022-05313-9}
  {\bibfield  {journal} {\bibinfo  {journal} {Nature}\ }\textbf {\bibinfo
  {volume} {611}},\ \bibinfo {pages} {260} (\bibinfo {year}
  {2022})}\BibitemShut {NoStop}%
\end{thebibliography}%

\end{document}